\documentclass[letterpaper]{article} 
\usepackage{aaai23}  
\usepackage{times}  
\usepackage{helvet}  
\usepackage{courier}  
\usepackage[hyphens]{url}  
\usepackage{graphicx} 
\usepackage{natbib}  
\usepackage{caption} 
\usepackage{algorithm}
\usepackage{algorithmic}

\usepackage{newfloat}
\usepackage{listings}
\DeclareCaptionStyle{ruled}{labelfont=normalfont,labelsep=colon,strut=off} 
\floatstyle{ruled}
\newfloat{listing}{tb}{lst}{}
\floatname{listing}{Listing}
\usepackage[utf8]{inputenc}
\usepackage{xcolor}
\usepackage{graphicx}
\usepackage{xcolor}
\usepackage{csvsimple}
\usepackage{textcomp}
\usepackage{amsmath}
\usepackage{multirow}
\usepackage{subcaption}
\usepackage{mwe}
\usepackage{chemformula}
\usepackage{url}
\usepackage{svg}

\newcommand{\equationtxt}[1]{Equation~#1}
\newcommand{\figuretxt}{Figure~}
\newcommand{\sectiontxt}{Section~}

\newcommand{\tabletxt}{Table~}
\newcommand{\eventbase}{Event-base Aggregation}
\newcommand{\timebase}{Time-base Aggregation}

\newcommand{\fscore}{F1-Score~}

\newcommand\keywords[1]{%
    \begingroup
    \let\and\\
    \par
    \noindent\textbf{Keywords:}\\#1\par
    \endgroup
}
\title{Detecting Security Patches via Behavioral Data in Code Repositories}
\author{
Nitzan Farhi \textsuperscript{\rm 1},
Noam Koenigstein  \textsuperscript{\rm 2},
Yuval Shavitt  \textsuperscript{\rm 1}
}

\affiliations{
\textsuperscript{\rm 1} School of Electrical Engineering, Tel-Aviv University, Tel-Aviv, Israel \\

\textsuperscript{\rm 2} Department  of Industrial Engineering, Tel-Aviv University, Tel-Aviv, Israel 

}
\date{May 2022}

\begin{document}

\maketitle

\begin{abstract}

The absolute majority of software today is developed collaboratively using collaborative version control tools such as Git. It is a common practice that once a vulnerability is detected and fixed, the developers behind the software issue a Common Vulnerabilities and Exposures or CVE record to alert the user community of the security hazard and urge them to integrate the security patch. 
However, some companies might not disclose their vulnerabilities and just update their repository. As a result, users are unaware of the vulnerability and may remain exposed.  

In this paper, we present a system to automatically identify security patches using only the developer behavior in the Git repository without analyzing the code itself or the remarks that accompanied the fix (commit message). We showed we can reveal concealed security patches with an accuracy of 88.3\% and F1 Score of 89.8\%.  This is the first time that a language-oblivious solution for this problem is presented.
\end{abstract}

\keywords{CVE, Machine Learning, Git, GitHub, LSTM, Conv1D}

\section{Introduction}
\label{sec:intro}

The absolute majority of software today is developed collaboratively, where developers work separately on a different part of the software, and then merge their code with the rest of the software. 
To allow this collaboration, source control tools, such as Git, were developed. Git is a standard protocol that supports file versioning, which is extensively used by programmers to collaborate on software development. A repository of files is maintained by GIT, allowing operations such as updating files (push/commit), merging file versions (merge), and splitting a project into two branches (fork). 

A hosting service such as GitHub\footnote{\url{https://github.com/}} allows users to maintain and share a repository, and also allows tracking issues such as bug reports, feature suggestions, fork repositories, maintain merge requests, etc. This makes GitHub an ecosystem of software, where interaction among developers reveal patterns of behavior~\cite{horawalavithana2019mentions}.  For example, adding a new feature to a repository may trigger many forks to the repository since other people will want to modify the feature.

Software is naturally prone to errors, and some of them can be exploited by hackers to abuse the software for malicious acts. These errors are often called vulnerabilities and are usually fixed by a software update called a security patch. Vulnerabilities can be divided into 2 classes: 0-day, a vulnerability that is not known to the public and still does not have a security patch, and N-day, one that has a security patch.

The Common Vulnerabilities and Exposures (CVE\footnote{\url{https://cve.mitre.org/}}) is a glossary of publicly disclosed vulnerabilities. This CVE glossary details the vulnerability date, description, references, and metrics regarding the vulnerability such as its complexity whether user interaction is required, and more.
Vulnerabilities are analyzed and the Common Vulnerability Scoring System (CVSS) is employed to evaluate the threat level. The CVE score is often used for prioritizing the security of vulnerabilities.

It is a common practice that once a vulnerability is detected and fixed,  the company or individuals behind the software issue a CVE to alert the user community of the security threat, hopefully urging them to integrate the security patch.
However, some companies might not disclose their vulnerabilities and just update their repository. This behavior may be in order to allow certain users to use the patch sooner, as a measure of ``Security by obscurity'', or to avoid public embarrassment. The disadvantage of not disclosing vulnerabilities is that users will not be aware that their software needs to be updated, while hackers can notice the code changes, detect a vulnerability, and exploit it.

Recent work \cite{sun2020identify, wang2021patchrnn} has shown the ability to analyze source code changes and commit messages using natural language processing models. These aforementioned models rely on a certain programming language or on the natural language used in the comments and commit messages. Hence, these earlier models are limited in their ability to detect security patches on a repository that uses other programming languages or where comments are written in a different natural language. 
For example, a model that was trained on detecting security patches in \textit{C++} would not be able to detect security patches in a \textit{Perl} repository, and since it is less popular, gathering a sufficient amount of data to train a model for \textit{Perl} would be difficult.

In this paper, we propose a method that trains a deep learning (DL) model to recognize the user behavioral data around a security patch of a repository on GitHub (a window of events that will be defined in the Data Pre-processing Section), and uses CVE publications to label if it is related to a security patch. Importantly, our model is insensitive to the programming language, or the natural language used by the developers. This allows us to identify security patches in languages that are rather rare and thus do not have sufficient data to train a model.
Our model was able to achieve an accuracy of 88.32\% and an \fscore of 89.75\% in detecting a security patch.  Hence, our approach can be effectively used for detecting undisclosed security patches.  

Since GitHub behavioral data is temporal (various actions over time), we experimented with appropriate models such as Long-Short Term Memory (LSTM) \cite{hochreiter1997long} and 1-Dimensional Convolutional Neural Network (Conv1D), which allow calculating relations between features at different time points.  
Note that we do not require Natural Language processing or manual labeling to detect security patches.  

We made the source code for our data extraction and our models publicly available at GitHub\footnote{\url{https://github.com/nitzanfarhi/SecurityPatchDetection}}. The relevant data-set is available as well\footnote{\url{https://nitzanfarhi.github.io/datasets}}.
We note that our published data-set covers more repositories than any previous work. 



\section{Related Work} \label{sec:related}
In this section, we provide an overview of different works in the field of detecting security patches using the CVE Program. 

PatchDB~\cite{wang2021patchdb} is a dataset of security patch commits on \textit{C/C++} projects. Each patch commit contains the additions and deletions made to the code and a commit message. This dataset was created by extracting security patch commits using the CVE Program (we use a similar extracting technique for our data collection).
PatchDB was used in PatchRNN~\cite{wang2021patchrnn} where natural language processing methods were employed to detect which commit is a security patch. This is done by tokenizing the commit message, the additions, and the deletions. Later, the tokens were fed into an LSTM neural network to provide a prediction regarding a specific commit with an accuracy of 83.57\%.  
A similar process was suggested by \citet{le2021deepcva}. A vulnerability-contributing commit can be deduced from the security patch (e.g., by using the command `git blame'). In \citet{le2021deepcva} tokenizing was performed on vulnerability-contributing commits in \textit{java} repositories and an attention-based GRU was employed in order to classify the severity of the commit. 

\citet{zhou2021spi} selected 4 open-source \textit{C/C++} projects and filtered commits by using known keywords that describe a security patch. Then, security researchers were employed to manually label the commits as either commits that \emph{introduces} a vulnerability, commits that \emph{closes} a vulnerability, or commits that are \emph{unrelated} to vulnerabilities. Natural Language Processing methods were used to create a deep learning model based on LSTM and Convolutional Neural network, which led to a reported accuracy of 90\%. 

\cite{li2017large} also used the CVE Program to extract security patches from git repositories. This was done to analyze and gain insights into the security patches. The authors investigated the number of changes done to the code base, the maturity of the bugs, and the number of bugs that were solved by the commits.

\cite{sun2020identify} extracted security patch commits from the CVE Program and verified that the commits are security patches by manually inspecting 200 of them. Then, the authors extracted the commit message only from GitHub (and did not use other features such as added lines, removed lines, or date). Repositories that contained C/C++ code were selected, and negative samples were acquired by taking all other commits from the same repositories. The natural language processing model that was used to detect the security patches was a hierarchical attention network with Gated Recurrent Units (GRU) \cite{cho2014learning} as its layers. \cite{sun2020identify} reached an accuracy of 92\%.

VCCFinder \cite{perl2015vccfinder} extract security patches of \textit{C/C++} repositories similarly, but used them to understand which is the commit that introduced the vulnerability. This paper takes the first step in the direction we suggest by extracting limited meta-data: they use present ``fork count'', ``star count'', ``number of commits'' and ``programming language'' but disregard the entire history of these event data. A Support Vector Machine (SVM) model was employed to detect the vulnerability introducing commits. 

\cite{pickerill2020phantom} used meta-data of Git logs to solve a different problem: detecting repositories that are ``engineered'', namely repositories that are not personal nor inactive. For this they extracted
Integration Frequency, Commit
Frequency, Integrator Frequency, Committer Frequency, and Merge Frequency, and  transformed it into a time series, which was used to detect whether a repository is ``engineered''.
Similarly, \cite{coelho2018identifying} used GitHub's API to extract the number of Forks, open issues closed issues, commits, and max days without commits to detect whether a project is under maintenance or unmaintained.

While there were some previous works that used the git behavioral meta-data, these works were either limited in the amount of meta-data that was used or focused on a different type of task.  This work is the first solution that uses meta-data alone, which enables a language-oblivious general solution. 

\section{Methods}
\label{sec:methods}

\subsection{Data Collection}
\label{sec:collection}
\begin{figure}
	\centering
	\includegraphics[width=0.5\textwidth]{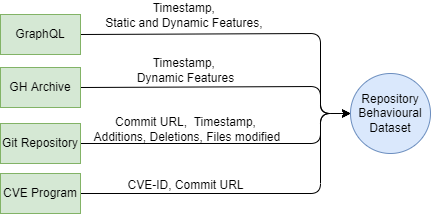}
	\caption{The overall data collection process.}
	\label{fig:data_collection}
\end{figure}

As explained below, we collected data from multiple sources and processed them into a single data-set that will be used by different models for training. A diagram of the overall data collection process can be seen in \figuretxt \ref{fig:data_collection}.

As mentioned before, Git is a collaborative version control system that allows tracking changes made to a repository. A \emph{commit} is a change to a repository that includes the text of the code that is being changed and a commit message describing the change in natural language.
To make the repository and the commits accessible to all developers, they can be sent (pushed) to GitHub, a server that mirrors the local Git information. 

GraphQL\footnote{\url{https://graphql.github.io/}} API allows us to iterate over the entire repository's history and gather all events. It must be noted that in this data collection, only data from ``master'', ``main'', or activate branches can be collected because inactive branches are deleted and cannot be extracted. The data was collected from the years 2015 to 2021 and features that were collected are detailed in \tabletxt\ref{tab:GitHub_features}.

We also used GitHub's GraphQL API to acquire static information about repositories, this information will allow us to create different models for different repositories (for example by programming language) and also improve accuracy, as will be detailed later. The static information includes properties of the repository such as the programming language in use, the current size of the repository, and information about the owner of the repository, e.g., if it is an individual or a company. The list of static features is detailed in \tabletxt\ref{tab:static_features}.

To acquire more data, we used ``GH Archive''\footnote{\url{https://www.gharchive.org/}}, which records GitHub events and can be easily accessed via SQL. This database contains events that cannot be acquired via GraphQL since they are recorded in real-time, and branches that were deleted from the repository cannot be retrieved. We acquired only the repositories that had CVEs and Gathered all the events from ``GraphQL'' and ``GH Archive'', making our dataset more accurate and comprehensive to detect unlabeled vulnerabilities. Features extracted from the GH Archive can also be seen in \tabletxt\ref{tab:GitHub_features}.

To gather a dataset of existing security patches, we used the largest existing database of vulnerabilities - the CVE Program. A CVE is a unique identifier that identifies a vulnerability, the CVE format is as follows: ``CVE-2021-44228'' where 2021 is the current year the vulnerability was published and 44228 is a unique vulnerability ID for that year. 

The CVE Program publishes all CVEs in a comma-separated values (CSV) format. Each entry in the CSV file is a CVE, and includes the CVE's  description and additional links. 
We used the additional links to acquire security patches in the following way: the additional links' field can be separated into separate links, and if one of the links points to a commit in GitHub, we gather the tuple \textlangle CVE-ID, commit-URL\textrangle.

Finally, we cloned the repositories from GitHub to acquire a mapping between a commit ID and its timestamp, the number of additions, deletions, and the number of files that were changed. This allows us to have timestamped labeled commit events, where a commit can be labeled as either \emph{positive} - namely, a security patch or \emph{negative} - not a security patch. Since all other data we acquired is also dated, we overall have a dataset of events with their date, time, and label.

\begin{table*}[]
	\begin{tabular}{|l|l|l|}
		\hline
		Feature                  & Type    & Explanation                                                \\ \hline
		isCompany                & Boolean & Does the repository belong to a company or an independent person \\ \hline
		isEmployee               & Boolean & Is the repository owner an employee of a company                 \\ \hline
		isHirable               & Boolean & Is the repository owner can be hired                             \\ \hline
		isSiteAdmin              & Boolean & Is the repository owner site administrator                       \\ \hline
		isSponsoringViewer       & Boolean & Is the repository owner sponsored                                \\ \hline
		isGitHubStar             & Boolean & Is the repository owner a member of the GitHub star program      \\ \hline
		isCampusExpert           & Boolean & Is the repository owner a member of the campus expert program    \\ \hline
		isDeveloperProgramMember & Boolean & Is the repository owner a member of the GitHub developer program \\ \hline
		isVerified               & Boolean & Is the repository owner verified by GitHub                       \\ \hline
		isInOrganization         & Boolean & Is the repository owner in an organization                       \\ \hline
		createdAt                & Integer & Year the repository was created (one-hot encoded)                \\ \hline
		diskUsage                & Integer & Disk usage of the repository                                     \\ \hline
		hasIssuesEnabled         & Boolean & Does the repository have issues enabled                          \\ \hline
		hasWikiEnabled           & Boolean & Does the repository have wiki enabled                            \\ \hline
		isMirror                 & Boolean & Is the repository a mirror                                       \\ \hline
		isSecurityPolicyEnabled  & Boolean & Is the security policy enabled in the repository                 \\ \hline
		fundingLinks             & Integer & Number of funding links in the repository                        \\ \hline
		languages                & List    & List of programming languages used in the repository             \\ \hline
	\end{tabular}
	\caption{Repository's static features that can be extracted via GraphQL.}
	\label{tab:static_features}
\end{table*}

\begin{table*}[]
	\begin{tabular}{|l|l|l|}
		\hline
		Feature                           & Description                                  & Location   \\ \hline
		Comment Event                     & Comment on a commit was added                & GraphQL    \\ \hline
		Stargazers Event                  & A stargazer was added to the repository      & GraphQL    \\ \hline
		Commit Additions Event            & lines were added to the repository (Integer)     & GraphQL    \\ \hline
		Commit Deletions Event            & lines were deleted from the repository (Integer) & GraphQL    \\ \hline
		Subscribers Event                 & A subscription was added to the repository   & GraphQL    \\ \hline

		Create Event                      & A Git branch or tag was created              & GH Archive \\ \hline
		Delete Event                      & A Git branch or tag was deleted              & GH Archive \\ \hline
		Issue Comment Event               & A comment on an issue was added              & GH Archive \\ \hline
		Commit Comment Event              & A comment on a commit was added              & GH Archive \\ \hline
		Pull Request Review Comment Event & A comment on a pull request was added        & GH Archive \\ \hline
		Member Event                      & Membership on the repository changed         & GH Archive \\ \hline
		Public Event                      & A repository changed its viewing settings    & GH Archive \\ \hline

		Push Event                        & A commit was pushed to the repository        & Both       \\ \hline
		Fork Event                        & The repository was forked                    & Both       \\ \hline
		Release Event                     & A new version of the repository was released & Both       \\ \hline
		Issue Event                       & A new issue was added                        & Both       \\ \hline
		Watchers Event                    & User marked the repository was watched       & Both       \\ \hline
		Pull Request Event                & A Pull request was created                   & Both       \\ \hline
	\end{tabular}
	\caption{Repository's features that can be extracted via GraphQL and GH Archive}
	\label{tab:GitHub_features}
\end{table*}

\subsection{Data Pre-processing}
\label{sec:preprocess}

The main preprocessing method we used is \eventbase:
Time elapsed between two events is not very meaningful, for example, it is insignificant if a fork was done one hour after a commit or two. Therefore, we can discard the date and time (and preserve only the temporal features that are one hot encoded).
A configurable parameter is the \textit{window\_size}, which denotes the number of events before and after the security patch commit that are concatenated together to create a window (or a vector) of features.
This window of features is one of many that will be fed into the model to train it, as detailed in the next section. 

Different from the time interval between events, other temporal features were found to be informative such as hour, day of the week, and month. These features, along with event types were encoded using one-hot encoding and fed to the LSTM units which usually take numbers and not events.
At this point, we gather for each repository all its security patch commits. Additionally, in order to make the dataset balanced, we also collected a random sample of non-security patch commits. 

For each repository, we created a list of events with their date and time. Commit events also have additional normalized integer features: the number of lines added to a file, the number of lines deleted from a file, and the number of files modified. Furthermore, repositories that contained less than 100 events overall were discarded, since they were mostly insufficient for extracting the windows that we required.

\subsection{Models}
\label{sec:lstm}
When dealing with large amounts of data, it is important to leverage the temporal information encapsulated in
the data. A few types of models were examined for this cause. LSTM is an architecture that had proven its efficiency for such tasks. LSTM is
based on Recurrent Neural Network (RNN) architecture, where performance decreases as a larger number of time steps are fed into the network.
However, LSTM can forget some less important data, while preserving the more important parts of it. GRUs have an architecture that is similar to LSTMs but with fewer parameters, which might be an advantage for some learning tasks. Finally, Conv1D is a Convolutional Neural Network (CNN) that convolutes 1-dimensional data, such as time series. This model can be stacked to allow multi-level pattern recognition. 

\begin{figure}
	\centering
        \includegraphics[width=0.5\textwidth]{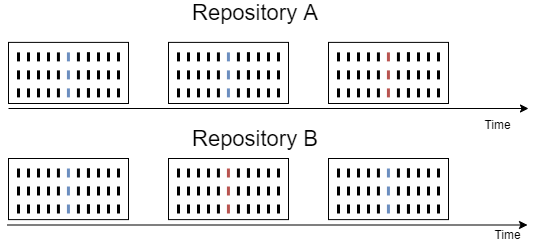}
	\caption{Demonstration of the sliding window method.}
	\label{fig:sliding_window}
\end{figure}

These model architectures accept their data as a series of vectors, each can be associated with many features. The
following process, as can be seen in \figuretxt\ref{fig:sliding_window} is employed to allow a time series to be given as input to LSTM, GRU, and Conv1D. For each repository, in our data-set, we gathered for each label - security and non-security patch commits, all the events that surround a commit, as a single input to the model. For example, an 11-event window (composed of a certain event, 5 events before and 5 events after) can be used to classify the window as one that contains a security patch (marked with a red column) or blue (as a window with non-security patches). As can be seen in the figure, some events might contain more than one feature (one column is composed of a few short lines).
Window size is the number of events that are contained in the window, each window size is constant, and windows are fed into the models one by one.

\section{Experimental Design}
\label{sec:experimental}
To validate the proposed methodology, the overall experimental design is as follows: 
We randomly divided the repositories into training (90\%) and test sets (10\%).
We further employed $k$-fold cross-validation on the training data to choose hyper-parameters, where $k=10$ as follows: On each iteration, 10\% of the repositories were kept as validation, which the model did not use for training.

The distribution that achieved the best validation accuracy was used for hyper-tuning the parameters, to achieve even better validation accuracy.
Finally, the obtained model was used to classify the windows that surrounds a commits of the test set, and performance metrics were calculated as described in \sectiontxt Performance Evaluation.

As mentioned above, we experimented with several learning models for training:
\begin{itemize}
    \item \textit{Conv1D} - uses Conv1D layer, Max Pooling, and then fully connected layers (\tabletxt\ref{tab:model-conv1d}).
    \item \textit{LSTM} - LSTM model, which uses LSTM neural network layers (\tabletxt\ref{tab:model-lstm}).
    \item \textit{GRU} - GRU model, which uses GRU neural network layers (\tabletxt\ref{tab:model-gru}). 
\end{itemize}

\begin{table}[] 
\begin{tabular}{lll} 
Model: Conv1D \\ \hline 
Type & Output Shape                 & Param \#     \\ \hline \hline 
Conv1D & (None, 198, 64)              & 8256        \\ \hline 
MaxPooling1d & (None, 99, 64)               & 0           \\ \hline 
Flatten & (None, 6336)                 & 0           \\ \hline 
Dense & (None, 256)                  & 1622272     \\ \hline 
Dropout & (None, 256)                  & 0           \\ \hline 
Dense & (None, 64)                   & 16448       \\ \hline 
Dropout & (None, 64)                   & 0           \\ \hline 
Dense & (None, 64)                   & 4160        \\ \hline 
Dropout & (None, 64)                   & 0           \\ \hline 
Dense & (None, 1)                    & 65          \\ \hline 
\hline 
Total params: 1,651,201 \\ 
\end{tabular} 
\caption{Model summary for Conv1D.} 
\label{tab:model-conv1d} 
\end{table}

\begin{table}[] 
\begin{tabular}{lll} 
Model: LSTM \\ \hline 
Type & Output Shape     & Param \#     \\ \hline \hline 
LSTM & (None, 199, 100) & 66000       \\ \hline 
LSTM & (None, 199, 50)  & 30200       \\ \hline 
LSTM & (None, 199, 25)  & 7600        \\ \hline 
LSTM & (None, 12)       & 1824        \\ \hline 
Dense & (None, 1)        & 13          \\ \hline 
\hline 
Total params: 105,637 \\ 
\end{tabular} 
\caption{Model summary for LSTM.} 
\label{tab:model-lstm} 
\end{table}

\begin{table}[] 
\begin{tabular}{lll} 
Model: GRU \\ \hline 
Type & Output Shape                 & Param \#     \\ \hline \hline 
GRU & (None, 199, 100)             & 49800       \\ \hline 
GRU & (None, 199, 50)              & 22800       \\ \hline 
GRU & (None, 199, 25)              & 5775        \\ \hline 
GRU & (None, 12)                   & 1404        \\ \hline 
Dense & (None, 1)                    & 13          \\ \hline 
\hline 
Total params: 79,792 \\ 
\end{tabular} 
\caption{Model summary for GRU.} 
\label{tab:model-gru} 
\end{table}

The proposed models were implemented using the Keras \cite{chollet2015} library with the TensorFlow backend
\cite{abadi2016tensorflow} in Python.
Training the model was done on a 64-bit Windows 10 Machine with Intel® Core™ i9-10940X CPU @ 3.30GHz with 16.0 GB RAM and NVIDIA GeForce RTX 3080 GPU, where training the training set takes about 4 minutes, and a single prediction takes about 5 milliseconds. The optimizer used is SGD and the selected batch size is 32  and the number of epochs is 50. 

\subsection{Performance Evaluation}
\label{sec:performance_evaluation}
To assess the performance of the trained models, we will define the metrics that were used.
When solving binary classification problems, accuracy, as defined in \equationtxt{\ref{equ:accuracy}}, is usually used as the main evaluation criterion.
Other important metrics are precision (\equationtxt{\ref{equ:precision}}), recall (\equationtxt{\ref{equ:recall}}), and \fscore (\equationtxt{\ref{equ:f1}}), which is calculated from precision and recall. These metrics are defined as follows: 
\begin{equation}
	\label{equ:accuracy}
	Accuracy = \frac{\textnormal{TP + TN}}{\textnormal{TP + FP + TN + FN}},
\end{equation}
\begin{equation} \label{equ:recall}
	Recall = \frac{\text{TP}}{\text{TP + FN} },
\end{equation}
\begin{equation} \label{equ:precision}
	Precision = \frac{\text{TP}}{\text{TP + FP} },
\end{equation}
\begin{equation} \label{equ:f1}
	F1 = 2 \cdot \frac{Precision \cdot Recall}{Precision + Recall},
\end{equation}
where $TP$, $TN$, $FP$, and $FN$ are the count of \emph{true-positives}, \emph{true-negatives}, \emph{false-positives}, and \emph{false-negatives} respectively. 
Finally, we also report the Receiver operating characteristic (ROC). The ROC is a curve that shows the trade-off between the false-positive rate and the true-positive rate according to a selected threshold. The area under the ROC curve (AUC) gives a numerical grade between 0 and 1 to the classifier, where 1 is a perfect classification.

\section{Results}
\label{sec:results}
\begin{figure}
	\centering
        \includegraphics[width=0.5\textwidth]{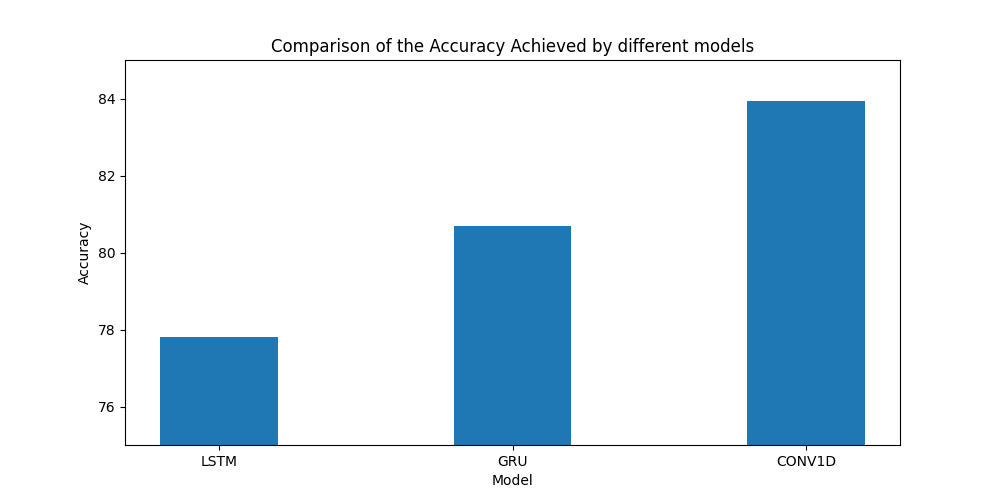}
	\caption{Comparison of the accuracy of different models.}
	\label{fig:model_compare}
\end{figure}

As discussed in \sectiontxt Experimental Design, a few models were evaluated to determine which is optimal for the task of classifying commits as security patches. 
We tested the three models with varied window sizes and found that for all window values Conv1D was the best model.  
%
%
\figuretxt \ref{fig:model_compare} compares the three model accuracy for the optimal window size of 10. Conv1D is leading by 3 percentage points over the next model and hence, in the rest of the experiments, we will focus on this model.

To discuss the rest of the results, we remind the reader that the window size determines the number of events that were gathered before and after the labeled event, e.g., if the window size is 5, we gathered overall 10 events: 5 before and 5 after the event).

\begin{figure}
	\centering
	\includegraphics[width=0.5\textwidth]{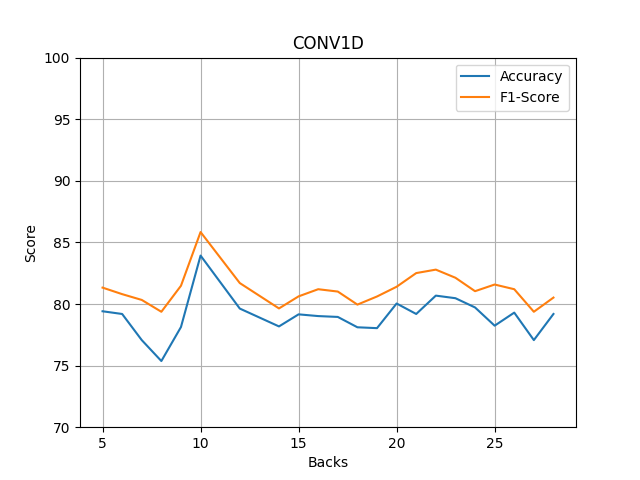}
	\caption{Conv1D model prediction accuracy and \fscore as a function of window size.}
	\label{fig:window_size}
\end{figure}

In the experiments, we checked the metrics when using different window sizes, which varied between 5 and 20. As can be seen in \figuretxt\ref{fig:window_size}, the optimal window size is 10 in both accuracy and \fscore, when using the Conv1D model.


\tabletxt\ref{tab:confusion-matrix} shows a confusion matrix for the best case, namely where the window size is set to 10 and the used model is Conv1D. In this case, the accuracy achieved was 83.93\%, \fscore was 84.14\%, the precision was 0.7550, and the recall was 0.8769. 

The number of false positives (FPs) is rather high, which results in rather low precision.  We suspect that some of the FPs are unreported security patches that our model identifies correctly.  To validate this we manually examined 52 randomly selected FPs and found (see also \sectiontxt Case Study) that 16 of them contain security patches, namely 30.8\%.  If we apply this to the confusion matrix of \tabletxt\ref{tab:confusion-matrix}, we get an estimate of 39 cases that are actually true positives (TPs). This results with a precision of 0.852. The accuracy also improves to 88.32\%, and the \fscore to 89.75\%.

\begin{figure}
	\centering
	\includegraphics[width=0.5\textwidth]{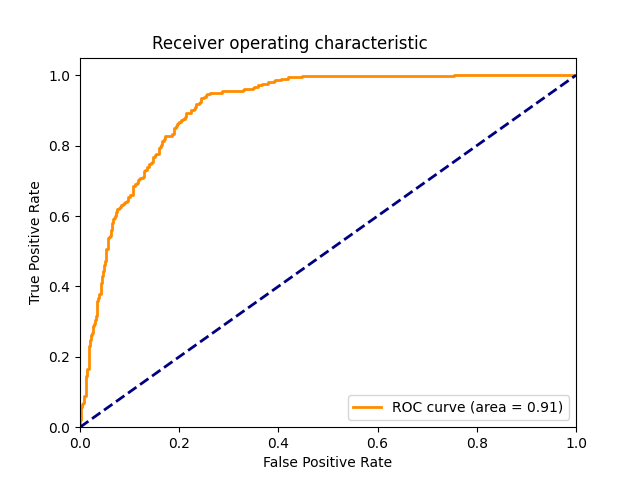}
	\caption{ROC curve for the best model.}
	\label{fig:roc}
\end{figure}

\figuretxt\ref{fig:roc} depicts the AUC for a window size of 10, which  is 0.91. 
Since we already discussed above that some of the FPs are actually TPs, the real AUC is higher.

\tabletxt\ref{tab:results-compare} compares our results with previous works that attempted to detect security patch commits. We noted that our paper is the only one that uses only behavioral git operations, and the amount of repositories used is much larger.

\begin{table}[]
	\begin{tabular}{|l|l|l|}
		\hline
		                                                                 &
		\begin{tabular}[c]{@{}l@{}}Not Classified\\  as a Security Patch\end{tabular} &
		\begin{tabular}[c]{@{}l@{}}Classified\\  as a Security Patch\end{tabular}       \\ \hline
		\begin{tabular}[c]{@{}l@{}}Not Marked as\\ a Security Patch\end{tabular}      &
		373 (TN)                                                         &
		128 (FP)                                                           \\ \hline
		\begin{tabular}[c]{@{}l@{}}Marked as\\ a Security Patch\end{tabular}          &
		28 (FN)                                                          &
		473 (TP)                                                           \\ \hline
	\end{tabular}
	\caption{Confusion Matrix of the classification of patches in the test set (140 repositories).}
	\label{tab:confusion-matrix}
\end{table}

\begin{table*}[]
\begin{tabular}{llll}
\hline
                             & Data Used                                  & Amount Of Repositories         & Accuracy Score \\ \hline
\citet{wang2021patchrnn}     & Commit message and code difference         & 313                            & 83.57\%        \\ \hline
\citet{sun2020identify}      & Commit messages                            & 993                            & 92.81\% \\ \hline
\textbf{Our Paper with original labels}           & Git \& GitHub behavioral data             & 1389                           & 83.93\%        \\ \hline
\textbf{Our Paper with estimated labels}           & Git \& GitHub behavioral data             & 1389                           & 88.32 \%        \\ \hline
\end{tabular}
\caption{Results Comparison.}
\label{tab:results-compare}
\end{table*}

\subsection{Other Experiments}

In this section, we report experiments that had not produced sufficiently positive results. 

\begin{itemize}
\item \timebase - As mentioned before, we used the \eventbase to build the horizon window. Another possibility is to use \timebase by counting the events by type during time frames (Hour, 2 Hours, Day, etc...) and summing them up to a single data row. This turned out to achieve a low accuracy rate since events can be sparse and result in many zero entries.
\begin{itemize}
\item \textit{Hours / Days} - if \timebase is used, the amount of hours/days recorded before the labeled event is the window size.
\item \textit{Resample} - if \timebase is used, we can aggregate all gathered events differently (for example, all the events in several hours can be aggregated into one vector)
\end{itemize}
\item Extracting location of repositories from GitHub - Since GitHub provides an API to the location of the owner we wanted to test if learning behavior by region or country can improve the prediction. Unfortunately, this field is free text and does not necessarily contain the actual location or is empty, which resulted in a high variety of answers and proved to be not useful.
\item Gathering a window before the actual security patch: this is an attempt to predict a security patch in the future or as it occurs, as some events can indicate preparation for a security patch. However, the amount of data before the security patch is not sufficient to make such a prediction, and data after is also needed to achieve satisfactory results. 

\end{itemize}

\section{Case Study}
\label{sec:case-study}
As discussed above, some vulnerabilities are fixed in commits but are not reported as vulnerabilities. In this case, our model will classify them as vulnerabilities, but they will not be labeled as such (false-positive), we manually examined the false-positive samples and checked for vulnerability-fixing changes in them, out of 52 false-positive commits, we found 16 to contain security patches (30\%). We will elaborate on two case studies.

\subsection{LibTIFF}
LibTIFF is a library for processing the TIFF image format. A Commit\footnote{\url{https://GitHub.com/vadz/libtiff/commit/5ed9fea523316c2f5cec4d393e4d5d671c2dbc33}} fixed 2 heap-based buffer overflow vulnerabilities and was detected by our model as a security patch, although it was not assigned a CVE (but only an issue at Bugzilla\footnote{\url{http://bugzilla.maptools.org/show_bug.cgi?id=2633}}, which we do not take labels from).
\subsection{Wireshark}
Wireshark is a network protocol analyzer that allows viewing network communication at several protocol layers. A Commit\footnote{\url{https://GitHub.com/wireshark/wireshark/commit/ba179a7ef7e660d788cbaade65982ffc7249b91f}} fixed a denial of service attack by null pointer dereference. This is clear from observing the code and also detected by our model but the commit was not assigned a CVE.
 
\section{Conclusion}
\label{sec:conclusion}
In this work, we proposed a methodology to gather repositories' behavioral meta-data from 3 different sources. After merging the different sources into one time series data-set, we used the CVE Program to label the data-set's commits to security patches and non-security patches. We introduced 3 types of time series models and discovered that Conv1D achieved the best classification accuracy of 88.32\%. We did not use natural language processing tools, and thus our results are language insensitive (both programming language and natural language).

Our results can be used to detect unreported security patches and warn the community to patch an open security problem. 
Our future research goal is to explore the impact of the different features on the classification, 
this can be done by using more interpretable models, comparing model accuracy on specific languages, etc.


\bibliography{main} 
\end{document}